\documentclass[aps,pra,twocolumn,superscriptaddress,showpacs,showkeys,amsmath,amssymb]{revtex4}

\usepackage{amsfonts}
\usepackage{amssymb,amsmath}
\usepackage{mathrsfs}
\usepackage{latexsym}
\usepackage{amsmath}
\usepackage[cp1251]{inputenc}
\usepackage{graphicx}
\usepackage{dcolumn}
\usepackage{bm}
\usepackage{color}

\begin{document}

    \title{Beyond mean-field properties of binary dipolar Bose mixtures at low temperatures}

    \author{Volodymyr~Pastukhov\footnote{e-mail: volodyapastukhov@gmail.com}}
    \affiliation{Department for Theoretical Physics, Ivan Franko National University of Lviv,\\
     12 Drahomanov Street, Lviv-5, 79005, Ukraine}

    \date{\today}

    \pacs{67.85.-d}

    \keywords{dipolar Bose mixtures, one-loop approximation, superfluid hydrodynamics}

    \begin{abstract}
    We rigorously analyze the low-temperature properties of
homogeneous three-dimensional two-component Bose mixture with
dipole-dipole interaction. For such a system the effective
hydrodynamic action that governs the behavior of low-energy
excitations is derived. The infrared structure of the exact single-particle
Green's functions is obtained in terms of macroscopic parameters,
namely the inverse compressibility and the superfluid density
matrices. Within one-loop approximation we calculate the
anisotropic superfluid and condensate densities and give the
beyond mean-field stability condition for the binary dipolar Bose
gas. A brief variational derivation of the coupled equations that
describe macroscopic hydrodynamics of the system in the external
non-uniform potential at zero temperature is presented.
    \end{abstract}

    \maketitle

\section{Introduction}
\label{sec1}
\setcounter{equation}{0}

During past decade the progress in creation of Bose-Einstein
condensates of atoms with large magnetic moments
\cite{Griesmaier,Lu,Aikawa} opened up a new area in the physics of
ultracold gases. The interplay of the anisotropic long-range
dipole-dipole interaction and a short-range repulsion leads to the
exciting properties of dipolar condensates
\cite{Baranov,Lahaye_Menotti,Baranov_Dalmonte}.

Interesting, but less studied is the behavior of dipolar Bose
mixtures. Depending on the system's parameters the phase diagram
of binary dipolar Bose condensates is characterized by the
alternation of stable-unstable regions \cite{Goral_Santos} and
complicated structures formation
\cite{Saito_Kawaguchi,Young-S_Adhikari}. The ground-state
properties of low-dimensional systems were also extensively
studied \cite{Gligoric_Maluckov,Xi_Li_Shi}. In the
quasi-two-dimensional case, where the presence of dipole-dipole
interaction leads to the maxon-roton spectrum \cite{Santos} of
elementary excitations, the two-component dipolar Bose system
exhibits phase separation caused by the softening of the roton
mode \cite{Wilson_Ticknor_Bohn_Timmermans}. The long-range
character of dipolar two-body interaction is responsible for the
creation of solitons with a large number of atoms
\cite{Adhikari_Young-S}. Even for the binary Bose condensate that
contains only one dipolar component a variety of non-trivial
phases emerges, including a novel quantized vortices phase
\cite{Shirley_Anderson} in the rotating system and robust
supersolid phases \cite{Wilson_Shirley_Natu} in the system on the
square lattice.

Recent observation \cite{Kadau} of the droplets formation in
dysprosium ($^{164}$Dy) condensate will surely stimulate further
studies of dipolar Bose systems. These experiments discovered an
intriguing phase of the system where the superfluidity is
accompanied by the broken translational invariance. Actually, they
revealed the role of quantum fluctuations
\cite{Wachtler_Santos,Saito} in the formation of new states of
matter.

In the present article by means of hydrodynamic approach we study
the properties of two-component dipolar Bose mixtures which, we
hope, will soon be implemented. In particular, it is indicated the
impact of the beyond mean-field effects on the stability and
superfluid properties of two-component dipolar bosonic gases.

\section{Formulation}
The considered model is characterized by the following action
\begin{eqnarray}\label{S}
S=\int dx \,\psi^*_a(x)\left\{\partial_{\tau}+\frac{\hbar^2 }{2m_a}\Delta+\mu_a\right\}\psi_a(x)\nonumber\\
-\frac{1}{2}\int dx\int dx'\Phi_{ab}(x-x')
|\psi_a(x)|^2|\psi_b(x')|^2,
\end{eqnarray}
here $x=(\tau, {\bf r})$ and the summation over repeated indices
$a,b=(A,B)$ is assumed. The first term describes two
non-interacting sorts of Bose particles with chemical potentials
$\mu_a$ and the second one takes into account both dipole-dipole
interaction as well as short-range repulsion between particles
$\Phi_{ab}(x)=\delta(\tau)\Phi_{ab}({\bf r})$
\begin{eqnarray}
\Phi_{ab}({\bf r})=g_{ab}\delta({\bf
    r})+\Phi^{(d)}_{ab}({\bf r}),
\end{eqnarray}
where $\Phi^{(d)}_{ab}({\bf r})=d_ad_b\frac{1-3z^2/r^2}{r^3}$,
i.e. all dipole moments $d_A, d_B$ are assumed to be oriented
along $z$ axis. We impose periodic boundary conditions with large
volume $V$ on the spatial dependence of complex fields $\psi_a(x)$
and with period $\beta=1/T$ ($T$ is the temperature) on
the imaginary time variable $\tau$.

In order to study the properties of low-energy collective modes we
pass to the phase-density representation of $\psi$-fields
\cite{Popov}
\begin{eqnarray}\label{psi}
\psi_a(x)=\sqrt{n_a(x)}e^{i\phi_a(x)}, \,
\psi_a^*(x)=\sqrt{n_a(x)}e^{-i\phi_a(x)}.
\end{eqnarray}
For the spatially uniform system we can use the following
decomposition of density and phase fields
\begin{eqnarray}\label{n_a}
&&n_a(x)=n_a+\frac{1}{\sqrt{\beta V}}\sum_{K}e^{iKx}n^a_{K},
\nonumber\\
&&\phi_a(x)=\frac{1}{\sqrt{\beta V}}\sum_{K}e^{iKx}\phi^a_{K},
\end{eqnarray}
treating $n_a$ as equilibrium densities of each component
\cite{Pastukhov_q2D}. We also define four-momentum $K=(\omega_k,
{\bf k})$ with condition ${\bf k}\neq 0$ imposed on the wave
vector, and introduce bosonic Matsubara frequency $\omega_k$.

\section{Structure of low-lying excitations}
The hydrodynamic approach allows us to proceed further
consideration in the canonical ensemble. The corresponding action
after substitution of $n_a(x)$ and $\phi_a(x)$ in Eq.~(\ref{S})
reads
\begin{eqnarray}\label{S_hydro}
S=S_{0}+S_{int}.
\end{eqnarray}
Here the first term is ($\delta_{ab}$ denotes the Kronecker
delta)
\begin{eqnarray}\label{S_0}
S_{0}={\rm const}-\frac{1}{2}\sum_{K}\left\{\omega_k\phi^a_K
n^a_{-K}-\omega_k\phi^a_{-K}n^a_{K}
\right.\nonumber\\
\left.+\frac{\hbar^2k^2n_a}{m_a}
|\phi^a_{K}|^2+\left[\frac{\hbar^2k^2}{4m_a n_a}\delta_{ab}
+\nu_{ab}({\bf k})\right]n^a_{K}n^b_{-K}\right\},
\end{eqnarray}
where $\nu_{ab}({\bf k})=g_{ab}+4\pi d_ad_b[k^2_z/k^2-1/3]$ is the
Fourier transform of two-body potentials; ${\rm const}=-\beta Vn_a
n_b\nu_{ab}(0)/2$ shifts the ground-state energy of the system and
$\nu_{ab}(0)=g_{ab}$ should be treated as a direction-averaged
value of $\nu_{ab}({\bf k})$ in the ${\bf k} \rightarrow {\bf 0}$
limit. From the latter fact we see that the mean-field
thermodynamics of the uniform dipolar Bose system is unaffected by
dipole-dipole interaction. It is easy to show by means of simple
diagonalization procedure that $S_0$ is the action of
non-interacting Bogoliubov quasiparticles with two branches of
excitation spectrum and the last term in Eq.~(\ref{S_hydro}) takes
into account interaction between them
\begin{eqnarray}\label{S_int}
S_{int}&=&\frac{1}{2\sqrt{\beta V}}\sum_{K,
Q}\frac{\hbar^2}{m_a}{\bf kq}\,n^a_{-K-Q}
\phi^a_{K}\phi^a_{Q}\nonumber\\
&+&\frac{1}{3!\sqrt{\beta
V}}\sum_{K+Q+P=0}\frac{\hbar^2}{8m_an^2_a}(k^2+q^2+p^2)
n^a_{K}n^a_{Q}n^a_{P}\nonumber\\
&-&\frac{1}{8\beta
V}\sum_{K,Q}\frac{\hbar^2}{2m_an^3_a}(k^2+q^2)n^a_{K}n^a_{-K}n^a_{Q}n^a_{-Q}.
\end{eqnarray}
These collisional terms are responsible for the simplest
quasiparticle decay processes providing the finiteness of the
elementary excitation lifetime.

Previously \cite{Pastukhov_InfraredStr} it was shown how to relate
parameters of low-lying excitations with measurable quantities of
the one-component dipolar superfluid. This analysis can be
naturally extended on the two-component Bose mixtures with the
dipole-dipole interaction. In particular, for all
diagrams with two external $n_K$-lines one obtains
\begin{eqnarray}\label{D_rhorho}
D^{ab}_{nn}(K\rightarrow 0)=\partial \mu_a/\partial n_b+4\pi
d_ad_b[k^2_z/k^2-1/3].
\end{eqnarray}
Mentioning that differentiation of every exact vertex function
with respect to $n_a$ adds one more zero-momentum $n^a_K$ line to
this vertex we conclude
\begin{eqnarray}\label{D_rho}
D^{abc}_{nnn}(0,0)=\frac{\partial^3 f} {\partial n_a\partial
n_b\partial n_c} =\frac{\partial^2\mu_a}{\partial n_b
\partial n_c},\ldots,
\end{eqnarray}
where $f$ is the free energy density of a two-component dipolar
Bose system. Note that only second-order vertices exhibit
anisotropic behavior in the long-wavelength limit, i.e., depending
on the angle between momentum and the direction of the external
magnetic field.

Similarly to the one-component case we can derive the second class
of identities supposing that each constituent of our system moves
with velocity ${\bf v}_a$, which is equivalent to the following
gauge transformation $\phi_{A(B)}(x)\rightarrow
\phi_{A(B)}(x)-m_{A(B)}{\bf rv}_{A(B)}/\hbar$ of the initial
action (\ref{S_hydro}). Rotational invariance in the transverse to
the external field plane ensures that thermodynamic quantities of
the moving Bose system depend only on ${\bf v}^{\perp}_a{\bf
v}^{\perp}_b$ and $v^{z}_av^{z}_b$, therefore for the free energy
density we have
\begin{eqnarray}\label{f_v}
f_{\bf v}=f+\frac{1}{2}\rho^{\perp}_{ab} {\bf v}^{\perp}_a{\bf
v}^{\perp}_b+\frac{1}{2}\rho^{z}_{ab}v^{z}_av^{z}_b+\ldots,
\end{eqnarray}
where introduced here symmetric matrices $\rho^{z}_{ab}$,
$\rho^{\perp}_{ab}$ have the meaning of superfluid mass densities
along and in the transverse direction to the dipole orientation,
respectively. In the zero-temperature limit the whole liquid is
superfluid and consequently the requirement of current
conservation leads to the following identities
\begin{eqnarray}\label{curr_cons}
\rho^{z}_{AA}+\rho^{z}_{AB}=\rho^{\perp}_{AA}+\rho^{\perp}_{AB}=m_A
n_A, \ \ (A\rightarrow B).
\end{eqnarray}
Taking into account expansion (\ref{f_v}) and fact that
differential operator $\frac{\hbar}{im_{A(B)}}{\bf k}
\frac{\partial}{\partial {\bf v}_{A(B)}}$ acting on the exact
vertex function raises the number of $\phi_{A(B)}(K)$-lines with
zero frequency and vanishingly small ${\bf k}$ we finally get
\begin{eqnarray}\label{D_phiphi}
D^{ab}_{\phi\phi}(K\rightarrow 0)=\frac{\hbar^2
k^2}{m_am_b}\left\{\rho^{\perp}_{ab}k_{\perp}^2/k^2+\rho^{z}_{ab}
k^2_z/k^2\right\},
\end{eqnarray}
\begin{eqnarray}\label{D_phirho}
D^{ab}_{\phi n}(K\rightarrow 0)=-D^{ba}_{n\phi}(K\rightarrow
0)=\delta_{ab}\omega_k.
\end{eqnarray}
In principle, absence of infrared divergences in the hydrodynamic
description permits to verify the last equality within
perturbation theory arguments. Moreover, the behavior of every
vertex function is qualitatively reproduced even on the one-loop
level. Finally, it should be noted that by using these two
differentiation rules one can obtain the long-wavelength
asymptotics of arbitrary exact vertex. Particularly, for
third-order vertices with two phase and one density lines we find
in the limit $K, Q \rightarrow 0$
\begin{eqnarray}\label{D_phi}
D^{abc}_{\phi\phi n}(K,Q)=-\frac{\hbar^2{\bf k}_{\perp}{\bf
q}_{\perp}}{m_am_b}\frac{\partial\rho^{\perp}_{ab}}{\partial
n_c}-\frac{\hbar^2 k_z q_z}
{m_am_b}\frac{\partial\rho^{z}_{ab}}{\partial n_c},
\end{eqnarray}
which together with (\ref{D_rho}) justify the effective Landau
hydrodynamic description of two-component dipolar Bose systems.

Equations (\ref{D_rhorho}), (\ref{D_phiphi}) and (\ref{D_phirho})
clearly demonstrate the phonon-like behavior of two branches of
excitation spectrum in the long-length limit and determine
direction-dependent sound velocities in terms of matrices
$\chi_{ab}({\bf k})=D^{ab}_{nn}(K\rightarrow 0)$, $\eta_{ab}({\bf
k})=D^{ab}_{\phi\phi}(K\rightarrow 0)/\hbar^2 k^2$
\begin{eqnarray}\label{det} c^4_{\bf
k}-c^2_{\bf k} \rm{Sp}\{\eta({\bf k}) \chi({\bf k})\}+\det
|\eta({\bf k}) \chi({\bf k})|=0.
\end{eqnarray}
Although we discuss the ground-state properties of the system, the
above equation is valid for all temperatures up to the
superfluidity transition point.

\section{Bose-Einstein condensation phenomenon}
Within our hydrodynamic approach it is easy to find out the
infrared structure of the one-particle spectrum. Therefore, the
main purpose of this section is to derive exact low-energy
asymptotic behavior of the normal
\begin{eqnarray}\label{psi*psi}
G_{ab}(x-x')=-\langle\psi_a(x)\psi^*_b(x')\rangle,
\end{eqnarray}
and anomalous
\begin{eqnarray}\label{psipsi}
\tilde{G}_{ab}(x-x')=-\langle\psi_a(x)\psi_b(x')\rangle,
\end{eqnarray}
Green's functions which are determined as statistically averaged
values of various pairs of $\psi$-fields. For the spacial
dimensionalities higher than two the above functions at equal
imaginary time arguments and at large particle spacing tend to the
constant
\begin{eqnarray}
G_{ab}(x)=\tilde{G}_{ab}(x)=-\sqrt{n_{0a}n_{0b}}, \ \ \tau=0, \,
r\rightarrow \infty,
\end{eqnarray}
where $n_{0a}$ is the Bose condensate density of sort $a$. The
number of particles in the lowers single-particle state is a
model-dependent quantity, which in the hydrodynamic approach can
be calculated as follows (see \cite{Pastukhov_InfraredStr} for
details)
\begin{eqnarray}\label{rho_0}
\sqrt{n_{0a}}=\lim_{\tau'\rightarrow \tau-0}\langle
\sqrt{n_a(x)}e^{i\phi_a(x')}\rangle{\big|}_{{\bf r}'={\bf r}},
\end{eqnarray}
or, equivalently $\lim_{\tau'\rightarrow \tau-0}\langle
e^{-i\phi_a(x)}\sqrt{n_a(x')}\rangle{\big|}_{{\bf r}'={\bf r}}$. Passing
to the four-momentum space
\begin{eqnarray}\label{G_ab}
\mathcal{G}_{ab}(P)=\int dx
e^{-iPx}{\big\{}\sqrt{n_{0a}n_{0b}}+G_{ab}(x){\big\}},
\end{eqnarray}
\begin{eqnarray}\label{tilde_G_ab}
\tilde{\mathcal{G}}_{ab}(P)=\int dx
e^{-iPx}{\big\{}\sqrt{n_{0a}n_{0b}}+\tilde{G}_{ab}(x){\big\}},
\end{eqnarray}
and taking into account equations (\ref{psi*psi}), (\ref{psipsi}),
(\ref{rho_0}) and our estimation for the infrared structure of the
hydrodynamic action we have
\begin{eqnarray}\label{G_ab_P}
\mathcal{G}_{ab}(P\rightarrow
0)=-\sqrt{n_{0a}n_{0b}}\langle\phi^a_P\phi^b_{-P}\rangle,
\end{eqnarray}
and $\tilde{\mathcal{G}}_{ab}(P\rightarrow
0)=-\mathcal{G}_{ab}(P\rightarrow 0)$. The applicability of the
above equation is not restricted to the low-temperature limit
where it generalizes the celebrated Gavoret-Nozi\`eres result
\cite{Gavoret_Nozieres,Pistolesi} on the mixture of Bose
particles, but it is also valid at finite temperatures in a
condensate region. In particular, a zero-frequency limit of this
formula is the extension of the Josephson result \cite{Josephson}
which due to presence of dipole-dipole interaction possesses
intriguing anisotropic behavior.

\section{One-loop calculations}
The absence of the infrared divergences guarantees that in the
weak-coupling limit properties of the system can be described
quantitatively even within the first-order perturbation theory. In
order to calculate anisotropic superfluid densities of the
two-component dipolar mixture we have to collect all diagrams with
two external phase lines. On the one-loop level the problem is
simple since the correction to every $D^{ab}_{\phi\phi}(K)$ vertex
in this approximation is given by two diagrams (see Fig.~1),
\begin{figure}
\centerline{\includegraphics
[width=0.35\textwidth,clip,angle=-0]{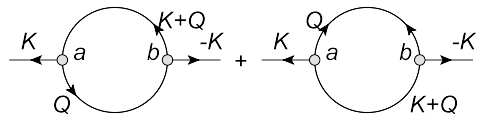}}
\caption{One-loop corrections to the $D^{ab}_{\phi\phi}(K)$
vertices. Arrows denote phase fields $\phi^a_K$ and solid lines
stand for density fields $\rho^a_K$.}
\end{figure}
so for the matrices of superfluid densities we obtain
\begin{eqnarray}
\rho^{z(\perp)}_{ab}=\delta_{ab}m_an_a
    -\Delta \rho^{z(\perp)}_{ab}.
\end{eqnarray}
At zero temperature these quantities satisfy the following
identities $\Delta \rho^{z(\perp)}_{AA}=\Delta
\rho^{z(\perp)}_{BB}= -\Delta \rho^{z(\perp)}_{AB}$ which are
consistent with current conservation (\ref{curr_cons}).
Straightforward calculations yield for $\Delta
\rho^{z(\perp)}_{AB}$
\begin{eqnarray}
\Delta \rho^{z}_{AB}=\frac{2}{
    V}\sum_{{\bf q} \neq 0}\hbar^2 q^2_z\frac{\varepsilon_A\varepsilon_B}{E_{+} E_{-}}\frac{n_A n_B \nu^2_{AB}}{(E_{+}+E_{-})^3},
\end{eqnarray}
\begin{eqnarray}
\Delta \rho^{\perp}_{AB}=\frac{1}{
    V}\sum_{{\bf q} \neq 0}\hbar^2(q^2_x+q^2_y)\frac{\varepsilon_A\varepsilon_B}{E_{+} E_{-}}
    \frac{n_A n_B\nu^2_{AB}}{(E_{+}+E_{-})^3},
\end{eqnarray}
where in order to simplify notation we introduced free-particle
dispersion $\varepsilon_a=\hbar^2q^2/2m_a$. We also used notation
for the two-body potential $\nu_{ab}=\nu_{ab}({\bf q})$ as well as
for two branches of excitation spectrum
$E^2_{\pm}=(E^2_A+E^2_B)/2\pm
\sqrt{(E^2_A-E^2_B)^2/4+4\varepsilon_A\varepsilon_B n_A
n_B\nu^2_{AB} }$, here
$E^2_A=\varepsilon^2_A+2\varepsilon_An_A\nu_{AA}$
($E_B=E_{A\rightarrow B}$) is the Bogoliubov spectrum of component
$A$ ($B$).

In the same fashion, by calculating one-loop diagrams contributing
to the $D^{ab}_{nn}(K)$ vertices (see Fig.~2)
\begin{figure}
    \centerline{\includegraphics
        [width=0.4\textwidth,clip,angle=-0]{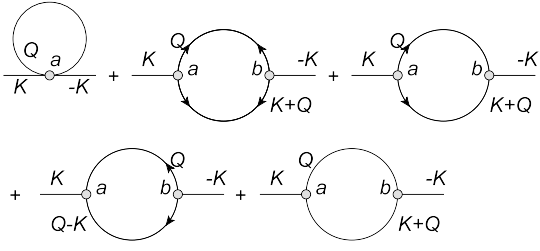}}
    \caption{Diagrams contributing to the $D^{ab}_{nn}(K)$ vertices.}
\end{figure}
we obtained corrections to the inverse susceptibilities
\begin{eqnarray}
\frac{\partial \mu_a}{\partial n_{b}}=g_{ab}+\frac{\partial
\Delta\mu_a}{\partial n_{b}}.
\end{eqnarray}
Omitting details of simple but cumbersome computations we present
results obtained in the zero-temperature limit
\begin{eqnarray}\label{Delta_mu_AA}
\frac{\partial \Delta\mu_A}{\partial n_{A}}=\frac{-1}{2
    V}\sum_{{\bf q} \neq 0}\frac{\varepsilon^2_A}{E^2_{+}E^2_{-}}\left\{
\frac{[\nu_{AA}E^2_B-2\varepsilon_Bn_B\nu^2_{AB}]^2}{E_{+}E_{-}(E_{+}+E_{-})}\right.\nonumber\\
\left.+\frac{\left[\nu_{AA}(E_{+}E_{-}+E^2_B)-2\varepsilon_Bn_B\nu^2_{AB}\right]^2}{(E_{+}+E_{-})^3}
\right\},
\end{eqnarray}
\begin{eqnarray}\label{Delta_mu_AB}
\frac{\partial \Delta\mu_A}{\partial n_{B}}=\frac{-1}{2
    V}\sum_{{\bf q} \neq 0}\frac{\varepsilon_A\varepsilon_B\nu^2_{AB}}{E^2_{+}E^2_{-}}\left\{
\frac{\varepsilon^2_A\varepsilon^2_B}{E_{+}E_{-}(E_{+}+E_{-})}\right.\nonumber\\
\left.+\frac{(E_{+}E_{-}+\varepsilon^2_A)(E_{+}E_{-}+\varepsilon^2_B)}{(E_{+}+E_{-})^3}
\right\}.
\end{eqnarray}
Of course, one obtains the same result using
thermodynamic identities $\frac{\partial \mu_a}{\partial
n_{b}}=\frac{\partial^2}{\partial n_{a}
\partial n_{b}}\frac{E_0}{V}$, where $E_0$ is the ground-state
energy of two-component dipolar Bose system in the Bogoliubov
approximation. It is worth noting that integrals in Eqs.
(\ref{Delta_mu_AA}), (\ref{Delta_mu_AB}) are ultraviolet
divergent. To remove these divergences we use the prescription
originally introduced in Ref.~\cite{Lima_Pelster} for the
one-component Bose system with dipolar interaction. The trick is
to rewrite the Fourier transform of potentials $\nu_{ab}({\bf k})$
via scattering amplitude at low momenta. Therefore, to the r.h.s.
of equations (\ref{Delta_mu_AA}) and (\ref{Delta_mu_AB}) we should
add $ \frac{m_A}{\hbar^2} \int\frac{d{\bf
q}}{(2\pi)^3}\nu^2_{AA}/q^2 $ and $
\frac{2m_Am_B}{\hbar^2(m_A+m_B)}\int\frac{d {\bf
q}}{(2\pi)^3}\nu^2_{AB}/q^2 $, respectively. The same result is
obtained with the help of usual dimensional regularization
procedure.

In the dilute limit the leading-order contribution to the
condensate depletion can be calculated in the closed form by
making use of Taylor series expansion of exponential and square
root factors in Eq.~(\ref{rho_0})
\begin{eqnarray}
\frac{n_{0A}}{n_{A}}&=&1-\langle \phi^2_A(x)\rangle
-\frac{1}{4n^2_A}\langle
n^2_A(x)\rangle\nonumber\\
&+&\frac{i}{n_A}\lim_{\tau'\rightarrow \tau-0}\langle \phi_A(x')
n_A(x)\rangle{\big|}_{{\bf r}'={\bf r}},
\end{eqnarray}
and similarly for the sort $B$. Performing integration over
Matsubara frequency for $\Delta n_{A}=n_{A}-n_{0A}$ we obtain
\begin{eqnarray}
\Delta n_{A}=\frac{1}{2
    V}\sum_{{\bf q} \neq 0}\left\{
    \frac{\varepsilon_A+n_A\nu_{AA}}{E_{+}+E_{-}}\left[1+\frac{E^2_B}{E_{+}E_{-}}\right]-1 \right\}.
\end{eqnarray}
The above result in the limit of vanishingly small interspecies
interaction coincides with that of Ref.~\cite{Schutzhold}.

For specific calculations we will restrict ourselves by
considering only mixtures with equal mass of different sorts of
particles. This is quite a reasonable approximation for
experimentally relevant two-component  $^{162}$Dy-$^{164}$Dy and
$^{164}$Dy-$^{162}$Er systems. Particularly, in this equal-mass
limit $m_A=m_B=m$ two branches of the excitation spectrum take a
very simple form $E^2_{\pm}=\hbar^4q^4/4m^2+\hbar^2c^2_{\pm}q^2$
providing that all thermodynamic parameters of a binary dipolar
Bose mixture at zero temperature can be written in terms of
one-dimensional integrals (see Appendix for details). Recent
experimental predictions \cite{Tang_Sykes,Maier_Ferrier-Barbut}
for the scattering lengths of dysprosium and erbium condensates
clearly show that the above mentioned homogeneous mixtures are
unstable towards collapse. Therefore, in order to demonstrate the
impact of the dipole-dipole interaction on the equilibrium
properties of two-component Bose gases we adopt simplified model
with $\det g_{ab}=0$ and $4\pi d^2_A/3g_{AA}=4\pi
d^2_B/3g_{BB}=\varepsilon<1$. This model is of particular interest
because such a system on the mean-field level is at the threshold
of phase separation and only the presence of quantum effects
recovers the thermodynamic stability of mixed state. Introducing
$s$-wave scattering lengthes $l_A=g_{AA}m/4\pi\hbar^2$
($l_B=l_{A\rightarrow B}$) and typical notation
$\mathcal{Q}_s(\varepsilon)=\int_{0}^{1}dx[1+\varepsilon(3x^2-1)]^{s/2}$
we are in position to write down condensate depletion
\begin{eqnarray}
\Delta n_{A}=
\frac{8}{3\sqrt{\pi}}n_Al_A\sqrt{n_Al_A+n_Bl_B}\mathcal{Q}_3(\varepsilon),
\end{eqnarray}
superfluid densities
\begin{eqnarray}
\Delta\rho^z_{AB}/m&=&
\frac{64}{45\sqrt{\pi}}\frac{n_Al_An_Bl_B}{\sqrt{n_Al_A+n_Bl_B}}\nonumber\\
&\times&\left\{\frac{1}{\varepsilon}\mathcal{Q}_5(\varepsilon)-\frac{1-\varepsilon}{\varepsilon}\mathcal{Q}_3(\varepsilon)\right\},
\end{eqnarray}
\begin{eqnarray}
\Delta\rho^{\perp}_{AB}/m&=&
\frac{32}{45\sqrt{\pi}}\frac{n_Al_An_Bl_B}{\sqrt{n_Al_A+n_Bl_B}}\nonumber\\
&\times&\left\{\frac{1+2\varepsilon}{\varepsilon}\mathcal{Q}_3(\varepsilon)-\frac{1}{\varepsilon}\mathcal{Q}_5(\varepsilon)\right\},
\end{eqnarray}
and corrections to the inverse compressibility matrix
\begin{eqnarray}\label{Delta_mu_AA_s}
\frac{1}{g_{AA}}\frac{\partial\Delta\mu_A}{\partial n_A}=
\frac{16}{\sqrt{\pi}}l_A\sqrt{n_Al_A+n_Bl_B}\mathcal{Q}_5(\varepsilon),
\end{eqnarray}
\begin{eqnarray}\label{Delta_mu_AB_s}
\frac{1}{g_{AB}}\frac{\partial\Delta\mu_A}{\partial n_B}=
\frac{16}{\sqrt{\pi}}\sqrt{l_Al_B}\sqrt{n_Al_A+n_Bl_B}\mathcal{Q}_5(\varepsilon).
\end{eqnarray}
As it is seen from the above calculations the mixture is always
stable for $l_A\neq l_B$. The sound velocity $c^2_{-}\simeq
n_An_B\det\chi({\bf k})/(mc_{+})^2$ of the lower branch of
excitation spectrum which is fully determined by the one-loop
result (\ref{Delta_mu_AA_s}), (\ref{Delta_mu_AB_s}) reveals
isotropic behavior and increases with increasing of the strength
of dipole-dipole interaction.

\section{Superfluid hydrodynamics}
In the following we consider only the low-temperature limit
completely ignoring densities of normal component in the
superfluid. To obtain the equations of macroscopic hydrodynamics
we use Zisel prescription adopted from
\cite{Griffin_Nikuni_Zaremba}. The underlying idea is the
formulation of variational principle that governs the evolution of
the local non-uniform densities $n_a({\bf r}, t)$ and velocities
fields ${\bf v}^{\perp}_a({\bf r}, t)$, $v^{z}_a({\bf r}, t)$ for
two-component dipolar bosons. The appropriate Lagrangian density
$\mathcal{L}=\mathcal{K}-\mathcal{U}$ necessarily contains kinetic
energy term (see Eq.~(\ref{f_v}))
$\mathcal{K}=\frac{1}{2}\rho^{\perp}_{ab}[n] {\bf
v}^{\perp}_a({\bf r}, t){\bf v}^{\perp}_b({\bf r},
t)+\frac{1}{2}\rho^{z}_{ab}[n]v^{z}_a({\bf r}, t)v^{z}_b({\bf r},
t)+\ldots$, where matrices $\rho^{\perp}_{ab}[n]$,
$\rho^{z}_{ab}[n]$ (through the local densities of each sort of
particles) depend on spatial coordinates and time. The second term
in $\mathcal{L}$ is the internal energy per volume of the
two-component Bose system with dipole-dipole interaction in the
external potentials $U_a({\bf r})$
\begin{eqnarray}\label{U}
\mathcal{U}&=&\epsilon[n]+U_a({\bf r})n_a({\bf r},
t)\nonumber\\
&+&\frac{1}{2}\int d{\bf r}'\Phi^{(d)}_{ab}({\bf r}'-{\bf
r})n_a({\bf r},t)n_b({\bf r}',t),
\end{eqnarray}
where $\epsilon[n]$ is energy density of the uniform system after
substitution $n_a\rightarrow n_a({\bf r},t)$. When minimizing
action $\mathcal{A}=\int dt \int d{\bf r} \mathcal{L}$ one should
take into account the local particle number conservation laws
\begin{eqnarray}\label{par_num_cons}
\partial_t n_a+\frac{1}{m_a} {\rm div}\,{\bf j}_a=0,
\end{eqnarray}
(from now on we do not write the dependence on ${\bf r},t$
explicitly) with currents defined by ${\bf
j}_a=(\rho^{\perp}_{ab}[n]{\bf v}^{\perp}_a,
\rho^{z}_{ab}[n]v^{z}_b)$. Performing these simple calculations we
obtain equations that determine conditional extremum of
$\mathcal{A}$
\begin{eqnarray}\label{v_t}
\partial_t{\bf v}_a=-\frac{1}{m_a}\nabla\left\{\mu_a[n]
+\int d{\bf r}'\Phi^{(d)}_{ab}({\bf r}-{\bf r}')n'_b\right.\nonumber\\
\left.+U_a({\bf r})+\frac{1}{2}\frac{\partial
\rho^{\perp}_{bc}[n]}{\partial n_a}{\bf v}^{\perp}_b{\bf
v}^{\perp}_c+\frac{1}{2}\frac{\partial \rho^{z}_{bc}[n]}{\partial
n_a} v^{z}_bv^{z}_c\right\},
\end{eqnarray}
which have to be solved together with (\ref{par_num_cons}). Few
comments are in order to outline the limits of applicability of the
above hydrodynamic equations. First of all, the total kinetic
energy $\mathcal{K}$ of two-component superfluid despite
one-component case is not a quadratic form over velocities even at
very low temperatures. Therefore omitting higher-order terms we
restrict our consideration to the distant hydrodynamic region
close to the equilibrium. In practice, using procedure of section
5 it is not hard to obtain these quadruple, etc. terms by
calculating appropriate vertices
$D^{abcd}_{\phi\phi\phi\phi}(K,Q,P)$, $\ldots$, but in the
experimentally relevant dilute limit their contribution are
negligibly small. Secondly, this semi-phenomenological formulation
of the macroscopic hydrodynamics is quasi-classical by its nature
which suggests the external potential $U_a({\bf r})$ to be a
smoothly varying function of coordinates, since only in this case
the so-called quantum pressure term $\hbar^2(\nabla
n_a)^2/(m_an^2_a)$ is smaller than chemical potential $\mu_a[n]$.
Moreover, in the absence of superfluid flow (${\bf v}_a=0$) from
(\ref{v_t}) we get the Thomas-Fermi stability condition for the
two-component system in the non-uniform external potential.
However, the obtained system of coupled equations can be easily
used to detect the formation of droplets with large number of
particles \cite{Wachtler_Santos} in mixtures with dipole-dipole
interaction. It is worth noting that the linearized equations
(\ref{par_num_cons}), (\ref{v_t}) correctly describe the
propagation of sound-waves in the homogeneous ($U_a({\bf r})=0$)
system with velocities given by Eq.~(\ref{det}).

\section{Conclusions}
To summarize, we have studied the properties of binary dipolar
Bose condensates at zero temperature. Making use of hydrodynamic
description in terms of density and phase fluctuations we have
found the connection between infrared anisotropic behavior of
one-particle Green's functions and dynamic structure factors with
macroscopic parameters of two-component superfluids. Within this
approach the matrices of superfluid densities and inverse
susceptibilities are calculated in the one-loop approximation for
a model of dipolar Bose gas with the short-range repulsion.

Additionally we pointed out on the correct way to calculate the
condensate density of interacting bosons in the Popov's
hydrodynamic formulation. The impact of the dipole-dipole
interaction on the condensate depletion in the two-component Bose
gas is examined.

Finally, using variational approach we have considered the problem
of macroscopic motion of two-component dipolar superfluids at very
low temperatures. The obtained system of coupled hydrodynamic
equations can be used to describe future experiments with binary
dipolar Bose condensates.

\begin{center}
{\bf Acknowledgements}
\end{center}

We thank Dr.~A.~Rovenchak for useful comments. This work was partly supported by Project FF-30F (No.~0116U001539) from the Ministry of Education and Science of Ukraine.

\section{Appendix}

In this section we present some technical details of our first
order perturbative calculations. In the case when masses of two
species of particles are equal to each other the inverse
susceptibilities of the binary dipolar Bose gas can be written via
integrals over polar angle
\begin{eqnarray}
\frac{\partial\Delta \mu_{A}}{\partial
n_B}=\frac{1}{3\pi^2}\int^{\pi}_0d\theta
\sin\theta\frac{mc^2_+}{n_An_B}\left(\frac{mc_+}{\hbar}\right)^3\nonumber\\
\times\gamma^4_{AB}f_1(\gamma),
\end{eqnarray}
\begin{eqnarray}
\frac{\partial\Delta \mu_{A}}{\partial
n_A}=\frac{1}{3\pi^2}\int^{\pi}_0d\theta
\sin\theta\frac{mc^2_+}{n^2_A}\left(\frac{mc_+}{\hbar}\right)^3\nonumber\\
\times\left\{ \gamma^4_{A}f_1(\gamma)
-2\gamma^2_{A}\gamma^2f_2(\gamma) -\gamma^4f_3(\gamma) \right\},
\end{eqnarray}
where in order to shorten notations we introduced following
functions
\begin{eqnarray*}
    f_1(\gamma)=\frac{3\gamma^4+9\gamma^3+11\gamma^2+9\gamma+3}{(1+\gamma)^3}, \nonumber\\
    f_2(\gamma)=\frac{\gamma^2+3\gamma+1}{(1+\gamma)^3}, \ \
    f_3(\gamma)=\frac{1}{(1+\gamma)^3}.
\end{eqnarray*}
We also use abbreviation for $\gamma=c_{-}/c_{+}$,
$\gamma_A=c_A/c_{+}$ and $\gamma^4_{AB}=n_An_B\nu^2_{AB}({\bf
k})/m^2c^4_+$. Here
$c^2_{\pm}=(c^2_A+c^2_B)/2\pm\sqrt{(c^2_A-c^2_B)^2/4+n_An_B\nu^2_{AB}({\bf
k})/m^2}$ are sound velocities of two branches of the excitation
spectrum of the system and $c^2_A=n_A\nu_{AA}({\bf k})/m$
($c_B=c_{A\rightarrow B}$) is the mean-field sound velocity of
each sort along. In the same fashion the matrix elements of
superfluid densities read
\begin{eqnarray}
\Delta \rho^{z}_{AB}/m=\frac{4}{15\pi^2}\int^{\pi}_0d\theta
\sin\theta
\cos^2\theta\left(\frac{mc_+}{\hbar}\right)^3\nonumber\\
\times \gamma^4_{AB} f_2(\gamma),
\end{eqnarray}
\begin{eqnarray}
\Delta \rho^{\perp}_{AB}/m=\frac{2}{15\pi^2}\int^{\pi}_0d\theta
\sin\theta
\sin^2\theta\left(\frac{mc_+}{\hbar}\right)^3\nonumber\\
\times\gamma^4_{AB} f_2(\gamma).
\end{eqnarray}
Finally, for the condensate depletion of sort $B$ we find
\begin{eqnarray}
  \Delta n_B=\frac{1}{6\pi^2}\int^{\pi}_0d\theta \sin\theta \left(\frac{mc_+}{\hbar}\right)^3\nonumber\\
\times\left\{\frac{1-\gamma^5}{1-\gamma^2}-\gamma^2_A\frac{1-\gamma^3}{1-\gamma^2}\right\}.
\end{eqnarray}

\end{document}